\documentclass[11pt]{article}
\pdfoutput=1

\usepackage[normalem]{ulem}
\usepackage[swedish,english]{babel}

\usepackage{color}
\usepackage{xcolor}
\usepackage{amssymb}
\usepackage{amsmath}
\usepackage{ae}
\usepackage{slashed}
\usepackage{graphics}
\usepackage{graphicx}
\usepackage{epstopdf}
\usepackage{url}
\usepackage{hyperref}

\usepackage{cite}
\usepackage{times}
\usepackage{xfrac}

\usepackage{fancyhdr}
\usepackage{enumerate}
\def\be{\begin{equation}}
\def\ee{\end{equation}}

\usepackage{tikz}
\begin{document}
\selectlanguage{english}
\frenchspacing
\pagenumbering{roman}
\begin{center}
\null\vspace{\stretch{1}}

{ \Large {\bf
Building flat space-time from information\\ exchange between quantum fluctuations}}
\\
\vspace{1cm}
Anna Karlsson$^{1,2}$
\vspace{1cm}

{\small $^{1}${\it
Institute for Advanced Study, School of Natural Sciences\\
1 Einstein Drive, Princeton, NJ 08540, USA}}
\vspace{0.5cm}

{\small $^{2}${\it
Division for Theoretical Physics, Department of Physics, \\
Chalmers University of Technology, 412 96 Gothenburg, Sweden}}
\vspace{1.6cm}
\end{center}

\begin{abstract}
We consider a hypothesis in which classical space-time emerges from information exchange (interactions) between quantum fluctuations in the gravity theory. In this picture, a line element would arise as a statistical average of how frequently particles interact, through an individual rate $dt\sim 1/f_t$ and spatially interconnecting rates $dl\sim c/f$. The question is if space-time can be modelled consistently in this way. The ansatz would be opposite to the standard treatment of space-time as insensitive to altered physics at event horizons (disrupted propagation of information) but by extension relate to the connection of space-time to entanglement (interactions) through the gauge/gravity duality. We make a first, rough analysis of the implications this type of quantization would have on the classical structure of flat space-time, and of what would be required of the interactions. Seeing no obvious reason for why the origin would be unrealistic, we comment on expected effects in the presence of curvature.
\end{abstract}

\vspace{\stretch{3}}
\thispagestyle{empty}
\newpage
\pagenumbering{arabic}
\tableofcontents
\noindent\hrulefill

\section{Introduction}
A motivation for making the following analysis is to consider an alternative scenario to when the Einstein field equations (EFE) in general relativity break down. The standard concept of space-time is that it constitutes a background, with definite singularities only at points of high curvature. At an event horizon, new physics is not required to be experienced by an infalling observer due to the EFE, but is a matter related to the information paradox \cite{Hawking:1976ra}, black hole complementarity \cite{Susskind:1993if,Stephens:1993an} and the firewall paradox \cite{Almheiri:2012rt}, where the concern is the entanglement structure between Hawking radiation and interior modes. The central issue \emph{might} be fundamentally different. In extending space-time past an event horizon by means of the equivalence principle, an important assumption is made: that space-time exists independently of information exchange. If this is false, space-time structure would undergo a change near event horizons, where information `output' (such as signals sent out) undergoes a change as the radial distance from the event horizon is decreased, and eventually is randomized in the sense that the only output is Hawking radiation. That is, if space-time were to be dependent on information exchange in each `volume element' $dVdt$, the communication in the radial direction (in relation to a black hole) could not be independent of the radial distance from the event horizon, since only Hawking radiation is emitted at the horizon. A change in information exchange (e.g. the rate of it) would mean an introduction of a new scale, i.e. altered physics at/near the horizon. To our knowledge, this type of scenario has not been analysed. Some models with altered physics in the radial direction of black holes that are likely to have similar effects include the fuzzball proposal \cite{Mathur:2005zp} and the quantum phase transition model in \cite{Chapline:2000en}.

A key question is if space-time can be modelled consistently as a phenomenon emergent from information exchange --- interactions --- between quantum particles, or if a background picture is required. While an interaction origin would be exotic, the persistent difficulties with quantizing gravity merits looking beyond standard considerations, e.g. by examining the assumption mentioned above. We look at what would define a scenario where space-time arises on a macroscopic scale in a hydrodynamic manner from interactions at the quantum level, in terms of the physics involved at the quantum and classical levels. Since this analysis concerns quantum \emph{space-time} rather than quantum \emph{gravity}, a first issue is flat space-time (as we will get to in \S\ref{s.mc}), which therefore is the focus of the present text. Within this setting, there is no evidence that an interaction scenario is more physical than a background, but neither do we find immediate reason to rule it out, leaving it a hypothesis of interest for further analysis.

A few reasons to look into a quantum interaction origin of space-time begins with that in focussing only on background models, relevant physics might be bypassed. Secondly, for space-time concepts (distance, relative angles, time evolution) to arise from interactions, the connections must concern entanglement\footnote{For example, an interaction origin of a scalar product would require pairwise spin 1/2 entanglement.}, as mentioned in \cite{Karlsson:2018tod}, and entanglement has been shown to be important to space-time \cite{VanRaamsdonk:2010pw} through the gauge/gravity duality \cite{Maldacena:1997re}. A space-time upheld by dynamic entanglement would be upheld by interactions, and a logical question is if that origin is restricted to the dual gauge theory. The presented hypothesis could represent a new take on how to model an entanglement origin, although we do not discuss entanglement structures here. Thirdly, some immediate effects of an underlying information exchange structure are supported by a standard phenomenon which is well-understood but paradoxal: wave-particle duality in particle diffraction. The two last points are suggestive at best, but still interesting.

\subsection{The type of emergence considered: hydrodynamic}\label{s.emh}
Emergence of space-time can occur in different ways (see \S\ref{s.comparison} for a comparison with some examples) but to allow for as rich a dynamics at the quantum level as possible we choose not to define that physics in relation to the observed space-time physics. Instead, we consider the quantum dynamics to be a separate theory, with space-time emerging in the sense of how hydrodynamics emerges from a microscopic theory. The main point with emergence in this sense is to not impose that quantum interactions be restrained to act in a certain way in relation to time as it appears in the classical regime, which would be the case if e.g. identical particles had to interact with a set frequency. Since classical time could be an emergent effect, it should not be a prerequisite. Supporting this scenario is the observed thermodynamic behaviour of space-time \cite{Jacobson:1995ab}, but also the varying quality of time, in that its relative rate in a subsystem depends on both system speed and proximity to massive objects. Such variation is typically associated with concepts that depend on other factors instead of being fundamentally defining in themselves.

In this type of emergence, an intricate theory existing at a small scale gives rise to a simpler theory at a larger scale in that details are averaged out in the transition from small to large scales. The average is what is observed at the larger scale. Hydrodynamics is such an averaged theory. In explaining what the scenario could be for space-time, we will make an analogy with temperature, which also is an effective theory of this kind and which should be easy to picture, in order to understand in what sense space-time is considered to emerge, rather than existing as a prerequisite background.

Consider the existence of a concept at a large scale (temperature or time). Regard it as an average of a theory at a smaller scale. At the smaller scale, individual particles have a type of `energy' (some intrinsic property) and a type of particle interaction that on average are compatible with the large scale concept and spatial gradients of it. However, at the small scale the energy of any given particle is not set by the averaged concept, nor is the result of an interaction uniquely determined by the gradient properties. Rather, from a large scale picture the energy value falls within some statistical distribution, as does the redistribution (of energy) from an interaction. In changing the viewpoint from the large to the small scale, the appearance of the dynamics changes drastically. There is no reference to the averages, instead the theory is that of the energy and the interactions redistributing it.

For temperature, these small scale entities are kinetic energy and particle collisions, at which the sum of the energy is conserved. That precise conservation gives rise to the existence of steady-state linear gradients on the large scale, and the expectation value of the kinetic energy (per particle) connects to temperature through the Boltzmann constant, e.g. with $\langle E_k\rangle=\frac{3}{2}k_B T$ in the kinetic theory of gases. Now, to claim that kinetic energy and its properties are defined or governed by the classic thermodynamic model of temperature would be misleading. To construct a model of kinetic energy exchange between particles based on temperature would be to lose detail as well. For example, by assuming the presence of temperature in an extended region, one also assumes that the interactions are regular throughout that region, i.e. not suppressed to the point where an emergent, `connected' temperature does not form, which would be the case if an insulating surface extended through the region. Insulation divides unconnected and separately equilibrating subsystems, and in the temperature model it needs to be assumed separately, and be put in by hand. In the particle interaction picture, however, insulation arises naturally in the limit where interactions are suppressed. In addition, it naturally leads to a disruption of a theory modelled on a region with only regular interactions, i.e. without insulation.

In considering a basis on interactions, a general model needs to consider hydrodynamic emergence, to account for scenarios where the interactions are absent, or not frequent enough to give rise to the averaged theory typically present with typical interaction rates. This is also true in our case, where we consider if space-time can arise from information exchange (i.e. interactions) at the quantum level. A possible scenario is just the same as for temperature, as described right above, only with time (or space properties) representing the large scale concept, instead of temperature. Possible interactions where theorized on in \cite{Karlsson:2018tod}, but those are not central to the present analysis. Importantly, a particle property corresponding to time \emph{might} be as indirectly connected to time as kinetic energy is to temperature, and to identify what the quantum properties really are is a challenge we will only touch upon in this article.

An example of possible `insulation' in space-time would be new physics at event horizons, as suggested in the firewall and fuzzball conjectures. Our current analysis aims at considering an interaction origin and the effects thereof, to see if such a scenario could be consistent, and if observable physics far away from event horizons support, contradict or is indifferent to the scenario.

\subsection{The model, at the quantum level}
At the quantum level, the basic suggestion is to replace the metric tensor of general relativity (in its role of giving rise to $ds^2$) with interactions between pairs of quantum particles, including (and heavily reliant on) quantum fluctuations. For a connection to the metric, we look to \emph{rates of information exchange}. The metric tensor can straightforwardly be reinterpreted as describing a connectivity between quantum particles in terms of a frequency, with length through $dl\sim c/f$, and relative passage of time for each quantum particle through its total rate of interaction, with $dt\sim1/f_t$. In a diagonal metric, we have $\langle c^2f_t^2\rangle= -g_{tt}$. However, as frequencies $(f_t,f)$ are merely dummy notations derived from the connection to the classical properties, not necessarily inherent to the quantum model (compare to $E_k$ and $T$). At the quantum level, $(f_t,f)$ can correspond to some particle energies (or internal properties) not directly related to time or space. One might e.g. consider a scenario where the $f_t$ of a particle contributes to time in the sense of how the kinetic energy of a particle contributes to the temperature in its surroundings; describing (in some sense) a likelihood of interacting.

An equilibrating process is assumed to exist, so that gradients are suppressed, and to begin with the points of reference can be thought of as on a lattice. We discuss this ansatz further in \S\ref{s.qnat}. Right here the argument is that an average over interactions such as these can give rise to an effective metric tensor while the quantum physics is allowed to be decidedly different from a background scenario.

The quantum interactions are the object of investigation in \S\ref{s.qnat}. We look at how they are restricted by properties of flat space-time (including information exchange effects at the classical level) and by the metric tensor. We give some comments on what to expect with curvature, and in the near-horizon region, but restrictions from and quantum effects associated with curvature require analyses beyond those given in the present text.

\subsection{The model, at the classical level}\label{s.mc}
An origin of space-time in information exchange (interactions) would have consequences at the classical level beyond what is described by general relativity, and not limited to event horizons or curvature. Note that while the emergent theory in classically safe regions (far away from not only high curvature, but even curvature giving rise to event horizons) must be compatible with general relativity, it is allowed to be more detailed, e.g. fitting within the diffeomorphism invariance of general relativity.

The simplest example of a direct consequence is in settings where flat space-time is a good approximation and general relativity would not distinguish between regions with vacuum vs a presence of obstacles to information flow. With vacuum, we here mean empty space-time where only quantum fluctuations are present, and the optionally present obstacles are extended objects made of matter\footnote{Matter is considered an obstacle to information flow since light does not propagate through dense regions of it in the same way as it propagates through regions of vacuum.}, not sufficiently massive to prevent flat space-time from be a good approximation on the scales considered. General relativity would not make a distinction between these two scenarios since it is a theory of curvature only. However, any model based on information exchange must depend on how information is transmitted through regions, and so a presence vs absence of obstacles to information flow must give qualitatively different results for the space-time configuration.

With information propagation restricted due to matter, the shortest path is around rather than across an obstacle. This property of information exchange implies non-trivial structure of the associated flat space-times, relative to the case without obstacles. We will call this \emph{relative geometry}, and model it on the connectivity of flat space-time, leaving $dt$ constant.

In specific, consider a region of space-time that is defined relative to a set of reference points\footnote{If the reference points are thought of as objects, these should be sufficiently far away not to have an impact on the space-time configuration in the considered region.}. Let the reference coordinate system be such that rays of light passing through describe straight lines when the region only contains vacuum. Now, if an obstacle describing a surface extending only through a fraction of the region is placed within the region, an information origin of space-time would mean a bending of the light rays around the surface edges, relative to the reference frame set by the vacuum behaviour. This would happen since the surface would describe an `insulating' surface (using the temperature analogy) in terms of the interactions present, which would be cut off in one space dimension along the surface (the object). At the edges, the shortest path for information flow would be around the edges, and that information exchange would reshape the space-time compared to its shape in the vacuum configuration. This bending can be also be compared with how heat flows past edges of insulating surfaces vs how it would flow in an absence of insulation, in the temperature analogy. The interaction origin is similar, but where temperature interactions are linear, those of space-time are not, as visible in the EFE, and in the gravitational force, with $\propto 1/r^2$.

Note that relative geometry refers to the space-time \emph{configuration}, and `relative' refers to that it only makes sense as a comparison between obstacle vs pure vacuum scenarios. In modelling relative geometry, we will use vacuum configurations as reference frames for the metric configurations, illustrating metrics that are flat, but different relative to a set of reference points. To describe the different space-times, it is necessary to discuss particle paths \emph{delineating} the space-time configurations. Light rays are frequently employed for that, but any particle path would be affected by a change in the space-time metric. The relative geometry is \emph{not defined by particle motion}. Particle motion is only an effect of the metric. The relative geometry is defined by the configuration of the present obstacles, and by the interaction properties at the quantum level. In this, light just happens to have two roles, first as one example of information flow (information exchange with massless properties) and secondly as light rays delineating the metric.

Since flat space-time is well-known, any observable effects should be well-known as well. Interestingly, our example of an insulating surface above has a direct parallel in e.g. a disc giving rise to particle diffraction, an example of the wave-particle duality. This type of effect is present for any configuration of apertures or edges. If the discussion above appeared too unspecific, simply replace `obstacles to information flow compatible with approximately flat space-time' with a tabletop experiment looking at the effects edges/apertures have on light rays, compared with how light propagates when those objects are removed. The difference is observable in relation to a set reference frame where light rays describe straight lines in the absence of the objects in question.

Among the observed classical, flat space-time physics, the wave-particle duality is a good candidate for an effect coinciding with that of relative geometry. It is also the only one. Because of this, we model relative geometry on the observed distortion of light rays near apertures and edges. The general effect of the wave-particle duality is in agreement with what relative geometry could give rise to, and warrants a closer look at the concept rather than the opposite. The object of interest is not a description of the diffraction phenomenon, but if an information origin could be plausible. The suggested scenario is not a pilot wave theory, but is rather similar to a path integral approach in that it relies on probabilities of path deviations dependent on the relative geometry. In our approach, a qualitative model of relative geometry in agreement with the wave-particle duality provides a way to assess the logic of an underlying structure of space-time based on information exchange, through what it implies for the quantum interactions.

Consequently, our analysis at the classical level is of space properties in flat space-time. The analysis connects to quantum interactions giving rise to space instead of time (the `insulating' surface is an aperture instead of a firewall) but is relevant as a possible indicator of a space-time origin in information exchange, which would impact both space and time. The reason for why we begin with analysing this effect (relative geometry), its consistency and subsequent restrictions on the quantum interactions, is because relative geometry constitutes a first deviation from the properties of a standard space-time scenario. Note that the concept fits within flat space-time. That an observer comparing one region with a reference frame based on the vacuum configuration might see non-trivial geometry is in agreement with diffeomorphism invariance.

We look at two concrete examples of relative geometry in \S\ref{s.relgeo}. These represent metrics that appear bent in the reference frame, but which represent flat space-time metrics. In each case, the pair of relative geometry metric and reference frame metric represent metrics general relativity does not distinguish between, where a space-time originating in information exchange would do so, i.e. in agreement with different physics being observed (in terms of particle paths) for the two distinct scenarios (obstacles vs not). The point made is that based on the symmetry of the particle diffraction, it is possible to identify a relative geometry metric that is compatible with general relativity, since it is flat. An illustration of the slit diffraction example can be found in figure \ref{emesh}. There, the reference frame $(x,y)$ is flat, and incident light rays would delineate straight lines in the absence of an aperture. With slits however, the maxima and minima of the light distribution after the aperture describe hyperbolae, which are bent with respect to $(x,y)$ but straight lines in the relative geometry delineated in figure \ref{emesh}. The lines in the figure describe lines of constant value of the variables $(\xi,\eta)$ in \eqref{eq.eflat} and \eqref{eq.eflatdel}. Hence, figure \ref{emesh} illustrates the relative geometry of a slit aperture, in a simplified example. Both $(x,y)$ and $(\xi,\eta)$ describe flat metrics compatible with general relativity, and are identical far away from the aperture. An information origin would give a physical preference to $(\xi,\eta)$, which is compatible with space-time giving rise to particle diffraction\footnote{The details of how particle diffraction might arise, including interference patterns and quanta of particle path deviations, are discussed at the beginning of \S\ref{s.flat}.}, but does not represent conclusive proof in any way. In summary, figure \ref{emesh} represents the answer to what different space-time configuration a passing particle would be subjected to in the presence of a slit, with an interaction origin of space-time vs not. Instead of the configuration indifferent to obstacles, ($x,y$), a coordinate system describing straight lines would be $(\xi,\eta)$. The notion of directions and angles imparted on passing particles would be that of the lines in figure \ref{emesh} instead of lines of constant value of $(x,y)$.

\subsection{The main assumptions and results}
In the ansatz we analyse, we make the following central assumptions on the physics of space-time:
\begin{enumerate}\itemsep0em
\item Space-time has an origin in information exchange.
\item Space-time emerges as an average from the quantum dynamics, in a hydrodynamic fashion.
\item An example of how varying information exchange rates (in space-time) effects particle paths can be found in the wave-particle duality. (This assumption is at the classical level.)
\end{enumerate}
In summary, the information exchange origin anzats is made because \emph{(i)} with it, there is a potential for a new scale at event horizons, and \emph{(ii)} connections between space-time and entanglement indicate information exchange to be relevant. Information exchange equals interactions, and to get to a general picture we need to (at least) consider the classical space-time to have arisen in a similar way to hydrodynamics. Space-time is assumed to exist as an average property, but the quantum interactions do not need to refer to space-time properties, and in case of a suppression of the interactions, the average does not need to display the properties it otherwise has. Supporting this consideration is the similarities of space-time with thermodynamics.

The main results are:
\begin{enumerate}\itemsep0em
\item A first consistency check of the ansatz: relative geometries compatible with general relativity can be modelled on particle diffraction.
\item Equilibrated quantum interactions must have a Gaussian fall-off to capture the relations of Euclidean geometry, which coincides with trivial particle diffraction for Gaussian apertures.
\end{enumerate}
Our analysis focusses on spatial properties rather than time directly, but connects to if space-time could have an origin in interactions, in terms of information exchange. The consistency check, which can be found in \S\ref{s.relgeo}, is at best indicative of the possibility of an information exchange origin. The partial result on the nature of the quantum interactions in \S\ref{s.qrest} is our only non-trivial result. It is in agreement with but does not rely on the assumptions made at the classical level. A Gaussian shape relates not only to a symmetry compatibility with Gaussian apertures, but to that consecutive interactions (and notions of angles/lengths) in flat space-time must be compatible with Pythagoras' theorem, as in \eqref{eq.abc}. In addition, we discuss quantum effects in terms of particle propagation at the beginning of \S\ref{s.flat}. 

In all, this initial analysis is limited by being qualitative and without consideration of curvature, but it reveals enough internal consistency for further analysis to be of interest.

\subsection{Outline}
Below, we begin with a rough analysis of the implications an interaction exchange origin would have on flat space-time, in \S\ref{s.flat}. This includes a qualitative outline of effects of relative geometry on particle paths, two examples of relative geometry in slit and edge diffraction, and some general comments on the effective (very classical) structure that relative geometry would represent. Here, \S\ref{s.relgeo} merely illustrates the concept whereas the first paragraphs in \S\ref{s.flat} and \S\ref{s.qnat} detail the physics of the hypothesis. In \S\ref{s.qnat} we describe the quantum interactions, as well as possible effects of curvature.

However, before the main analysis we give a more detailed explanation of the type of emergence considered in the present text, and how it relates to other approaches to emergent space-time.

\subsection{Comparison with different versions of `emergence' of space-time}\label{s.comparison}
There exist many approaches to emergence of space-time from different `fundamental entities'. In connection to information theory there are models using the gauge/gravity duality by entanglement entropy \cite{VanRaamsdonk:2010pw} such as tensor networks \cite{Swingle:2009bg,Swingle:2012wq}, where emergence occurs in the sense of said duality, and the building block is entanglement between particles. Another type of approach includes direct discretization of space and time, as in causal set theory \cite{Bombelli:1987aa}, dynamical causal triangulation \cite{Ambjorn:2004qm} and loop quantum gravity \cite{Smolin:2004sx,Ashtekar:2004vs}. In these examples, emergence refers to how the continuum arises from a discrete theory. The present analysis is \emph{not} of either kind.

We consider a scenario where an effective theory arises at and above the Planck scale from more complicated dynamics at smaller scales, in the sense typically associated with hydrodynamics, hence the temperature likeness above. The building block under analysis is interactions by and between particles (quantum fluctuations), treated as events in the gravity theory directly. The central question is of an information theoretical nature, and is inspired by the observed similarity and likely connection of space-time to thermodynamics \cite{Jacobson:1995ab} and information theory.

Note that this type of effective theory must contain general relativity, but is not defined by curvature alone. Despite that the interesting physics of space-time often is defined as curvature, the governing properties of space-time also extend to how particles propagate through regions of (approximately) flat space-time. Properties of space must be part of the effective theory, and are indicators of the underlying structure as well. This is key in a scenario where the thermodynamic and information theoretic connections are tested in terms of a hydrodynamic emergence. With an emergence from interactions, i.e. with a theory built on information exchange, approximately flat space would display what we term `relative geometry' \emph{in relation to} obstacles to information flow (without obstacles, the term is not defined, it is relative only). That is, particle paths would change between scenarios with/without a certain obstacle present, due to a change in information flow. To understand whether or not this type of particle propagation, seen in the wave-particle duality, is connected to actual space properties (in effect, space-time properties) is relevant for understanding a possible sub-Planckian theory. This is the issue we raise and describe, and we conclude that a connection cannot be easily ruled out: it is relevant to discuss what role an information exchange origin vs particle diffraction may have in space-time physics.

A reader's first question might be how interaction frequencies $(f,f_t)$ and a light velocity $c$ as mentioned above can be defined without pre-existing notions of time and space. Classically, an event is defined by its position in space-time, but the definition is two-wayed. In an emergent scenario, one has the option of a breakdown of the dual process, where one can consider the scenario where the existence of multiple, ordered events on average gives the effect of time and space as we know it from classical physics. In this sense, the space-time-related qualities of $(f_t,f/c)$ do not define the interactions, instead they appear in an average of the interaction processes, as effective coefficients arising from a sublevel theory. We use the frequency notation only to relate the interactions to the classical properties. In the effective theory, the details of how effective coefficients arise are irrelevant: their existence and value is what is observed. This is typical of hydrodynamic scenarios.

Recall the analogy with temperature in \S\ref{s.emh} and the scenario where individual interactions in the form of events (with individual fluctuations) create an average interconnecting structure with averaged agreed upon individual events and interconnecting events, which in the effective theory translate into (e.g.) $\langle f_t\rangle$ and $\langle f/c\rangle$ respectively. From this angle, $(f_t,f/c)$ connect to how space and time appear \emph{due to the interactions}, contrary to the standard notion that space-time by its existence defines how events in it occur. The theoretical overlap is only at the level of the effective theory. In this way it is also possible to take a step further in independence of background assumptions; one can work independently of standard notions of time and space, as they are not taken to be defining features of the sub-Planckian theory.

The objective of the present analysis is to investigate if the consequences of such an information exchange origin are realistic or not; not to formulate a sub-Planckian theory of such interactions. The hypothesis has effects in terms of \emph{(i)} geometry\footnote{The relative geometry discussed in \S\ref{s.relgeo} might possibly be related to the model of geometry emerging from quantum bits suggested in \cite{Trugenberger:2016viw,Kelly:2019rpx}.}, through non-trivial space-time effects on geodesics apart from curvature, and \emph{(ii)} an additional scale for the breakdown of space-time, due to reliance on information exchange. The additional scale occurs since the quantum space-time theory would need to equilibrate in terms of information exchange, connected to $c$, in a feedback system.

\section{Flat space-time effects}\label{s.flat}
In setting up a first, crude model of relative geometry, suitable for a comparison with light diffraction, it is reasonable to restrict to static, flat space-time with geometry obstructing information flow surrounded by vacuum. $dt$ is constant. The scale of the geometry (introduced by various obstacles) as well as any other relevant physics is assumed to be such that no quantum effects of fluctuations belonging to the space-time structure are relevant to leading order. The stress-energy tensor is disregarded (particles passing through are assumed to have a negligible effect on the configuration) as well as any cosmological constant. Without relative geometry, the setting is trivial.

With relative geometry, two points of reference are used: the particle paths in the absence of obstacles (defining straight lines in a vacuum reference frame), and the obstacles setting the geometry. A flow of information is assumed to delineate geodesics through or past apertures or edges. In this sense the Minkowski space appears in a very special set of coordinates compared to the reference frame, and is dependent on the symmetries of the geometry. As is discussed in more detail in \S\ref{s.infoflow}, an intuitive origin would lie in how information spreads out from a constriction and in that sense sets a connectivity of a tension structure that would make up space-time. How tightly the tension structure is bound together would describe how quickly information passes through different regions. Effectively, the flat frame is a boundary value problem.

More than the configuration (apart from if relative geometries can be fit within flat space-time) the quantum effects of how particles pass through the space-time are of central importance. Unlike in a background, with an information origin of space-time a particle cannot simply `pass through space-time', following geodesics set by curvature. A quantum particle would interact with the quantum constituents underlying the tension structure which defines e.g. what is `forward'. This must be imparted on the particle, which cannot have knowledge of the local structure in each region prior to encountering it. Any such communication will be imperfect, and restricted by the way the particle interacts with the structure.

The process of particle propagation must be modelled on the wave-particle duality, which provides the only flat space-time effect that can be consistent with a relative geometry scenario. The key properties of such particle diffraction is scale invariance in particle wavelength $(\lambda)$ vs scale of the geometry ($D$), interference, end particle paths of straight lines in accord with the symmetry of the geometry, and nested intensity patterns from nested geometry configurations. A naive protocol can be outlined in terms of a series of steps, where a particle samples the local tension structure of a region or along a line of $O(\lambda)$ in comparison with its previous reference frame, gets an estimate of how the path bends $(\sim D)$, generates a shift in angle of progress according to some probability distribution and then reset its reference frame in agreement with the last sampled region. Here, the only relevant scale would be $\lambda$ vs how sharply the space bends over the sampled region, giving an effective scale invariance in $\lambda/D$. In the absence of curvature, the probability distribution for how angle deviations are generated need to be assumed to fall off quickly away from zero, so that particles with $\lambda\gg D$ remain insensitive to the structure. In addition, a reasonable quantum feature is a preference for certain quantum steps, here in terms of angular deviations instead of energy levels, creating periodic resonances within the diffraction intensity distributions.

The above picture may well be too simplistic to reproduce particle diffraction faithfully, but the principle is made clear. With a reference structure deviating from the vacuum configuration and in accord with the flow of information through the geometry, it is possible to obtain the properties of particle diffraction through a protocol similar to a path integral formalism, with particle wavelength $\lambda$ setting sensitivity to the reference structure\footnote{Effects dependent on $\lambda$ would have to arise from interactions between the particle and the reference structure.} --- at least provided that interference effects can be accommodated. We see no immediate objection to this (especially as preference for specific quanta is a characteristic of quantum physics) but a more thorough analysis would be required to fully verify that the correct diffraction can be captured. If not, the hypothesis of an interaction origin clearly fails.

Another concern is light diffraction with selection on polarization. We briefly comment on how this might be fitted into the picture after analysing the implications for the quantum interactions in \S\ref{s.qrest}. With a scenario building on information exchange, however, properties directly connected to information propagation (such as light) turn into defining qualities (in this case of space-time) rather than limitations on what is physical. Therefore polarization concerns, relating to characteristics of information without impact on most particles, are not among the most pressing matters at hand, and are deferred until a later point.

Lastly, the present analysis is of static relative geometries for simplicity, and generic relative geometries must include time dependence. With changes in geometry consistent with approximately flat space-time, causality should limit the propagation of the subsequent effects (classical or quantum).

We now proceed with illustrating two types of relative geometries that are central to diffraction patterns, the case of a single/double slit and that of a sharp edge, to illustrate the concept of relative geometry. It is classical in the same sense that general relativity is, with the difference that there is structure from information connectivity in the absence of curvature. The actual quantum constituents and their interactions, giving rise to the effective tension structure, is another matter entirely. In discussing relative geometry, we have merely assumed such a quantum level structure based on information exchange \emph{to exist}, and restricted the discussion to a region and scale where it is valid to disregard them in every sense except for their average impact. An interesting feature is that quantum effects must be present anyway, through particle interaction with the relative geometry.

\subsection{Relative geometry in flat space-time}\label{s.relgeo}
As described above, relative geometry would be present already in flat space-time and occur due to obstacles to information flow. It would introduce extra structure at changes in information connectivity, and so be non-trivial near e.g. apertures and edges. For a pair of simple key illustrations in static, flat space-time with constant $dt$, we restrict to $2d$ slices,
\be\label{eq.s2}
ds^2=-dt^2+dz^2+dl^2_{d=2}\,,\qquad dl^2_{d=2}=g_{ij}({\bf x})dx^idx^j\,,\quad dl^2_{d=2}\big|_\text{vacuum}=dx^2+dy^2\,,
\ee
with $g_{ij}({\bf x})$ smooth and where the added geometry is assumed to be constant in $z$ for a sufficient distance, such as for a long slit. The condition for the Riemann curvature tensor to vanish e.g. is
\be \label{2flat}
dl^2_{d=2}=f^2(u,v)du^2+h^2(u,v)dv^2\,:\quad \left(\frac{f'_v}{h}\right)_v'+\left(\frac{h'_u}{f}\right)_u'=0\,,
\ee
where $f'_v$ denotes a partial derivative: $f'_v=\partial f/\partial v$ etc. Relative geometry simply refers to a flat space-time which does not have a metric equal to the identity matrix in the vacuum reference frame $(x,y)$. Away from obstacles, the relative geometry needs to approach the reference frame quickly, likely exponentially fast.

The key to a relative geometry is the symmetry of the set-up, and for that we now proceed with two illustrations: slit and edge diffraction. Afterwards, we discuss an information connectivity interpretation of the concept.

\subsubsection{Single- and double-slit diffraction}
In the case of single- and double-slit diffraction, the symmetry of the set-up is directly observable in the lines delineating maxima and minima of the probability distribution of photons after diffraction by the slit in question. In a relative geometry setting, these lines describe how particles on average are guided through space, setting a notion of what straight propagation means on a particle level, in relation to the apertures, which are obstacles to information exchange. The overall pattern described (not focussing on details near the slits) is one of hyperbolae, since the maxima and/or minima follow lines of constant length difference to the two foci, embodied by either the edges of the one slit, or by the two slits.

Since the wave pattern describes hyperbolae for single or double slits, the symmetry of the light diffraction is that of an elliptic cylindrical coordinate system. In that setting, the hyperbolae describe lines parameterized by one of the coordinate variables, providing an elementary example of straight lines. A large part of the appropriate geometry hence is elliptic, and we will illustrate that the elliptic geometry is compatible with a flat space-time metric that seamlessly connects to the vacuum flat metric far away from the apertures.

The full symmetry picture is quite intricate, since the configuration of the obstacles have nested geometrical qualities. For a double-slit, there are single slits at the foci, and for a single slit, there are edges at the foci. In addition, an elliptic transformation does not give a diffeomorphism for the line between the foci. However, we will assume a decent separation of scales between the various effects, so that the different structures can be considered to be nested within each other and dealt with separately. We begin with the elliptic symmetry, and come back to the validity of the overall configuration afterwards.

\begin{figure}[tbp]
\begin{center}
\includegraphics[scale=0.7]{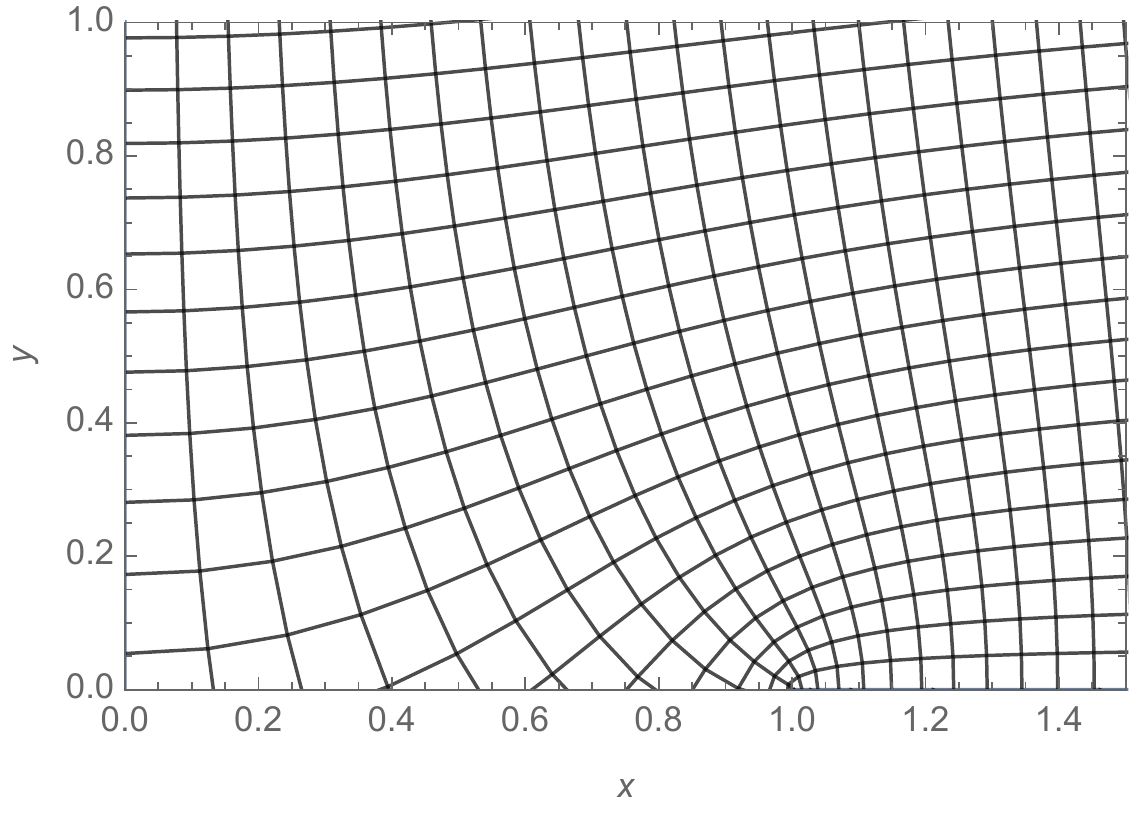}
\vspace{-0.5cm}
\end{center}
\caption{$2d$ illustration of how flat space-time can support `relative geometry' relative to a vacuum reference frame. Boundary conditions around geometrical objects (apertures, edges etc.) may cause local coordinate redefinitions of the actual flat frame. The mesh delineates a (2d cut-out of) flat space-time that causes hyperbolic geodesics in the reference frame. An illustration of those geodesics can be found in figure \ref{dofs}. Further adjustments need to be made along the slit opening ($|x|<1,y=0)$ for smoothness and edge effects.}\label{emesh}
\end{figure}

With the slit(s) extending in the $xz$-plane, symmetrically around $x=0$, we let the elliptical coordinates $(a,\phi)$ be
\be
\left\{\begin{array}{l}x=L \cosh a\sin\phi\\ y=L\sinh a \cos \phi\end{array}\right.\,, \qquad a\geq0\,,\quad \phi\in[0,2\pi]\,,
\ee
and from now on restrict to $2d$. $D=2L$ defines the distance between the foci, and we have
\be \label{vacell}
dx^2+dy^2=L^2\frac{\cosh2a+\cos2\phi}{2}(da^2+d\phi^2)\,.
\ee
In the elliptic coordinate system, the hyperbolae describe geodesics\footnote{$\ddot x^i+\Gamma^i_{jk}\dot x^j\dot x^k=0$ with only $\dot a\neq0$ gives $\dot a(t)=Ce^{-a(t)}$, with some constant $C$.} for
\be \label{ellflat}
dl^2_{d=2}=L^2\frac{e^{2a}}{4}\left(da^2+d\phi^2\right)\,.
\ee
This is {\it flat and has the reference frame as an asymptote} at $a>>1$. \eqref{ellflat} obeys \eqref{2flat} and approaches \eqref{vacell} exponentially fast in $a$. It can be recast into
\be\label{eq.eflat}
\eqref{ellflat} = \left[\begin{array}{c}\xi=e^a\sin\phi\times L/2\\\eta=e^a\cos\phi\times L/2\end{array}\right]=d\xi^2+d\eta^2\,,\qquad\xi^2+\eta^2\geq \frac{L^2}{4} \,,
\ee
with
\be\label{eq.eflatdel}
\left\{\begin{array}{l}x=\left(1+\frac{L^2/4}{\xi^2+\eta^2}\right)\xi\\y=\left(1-\frac{L^2/4}{\xi^2+\eta^2}\right)\eta\end{array}\right.\,.
\ee
The lines of constant $(\xi,\eta)$ are depicted in the $xy$-plane in figure \ref{emesh}.

The upshot of the above is that through the symmetry, it is possible to identify a flat geometry where hyperbolae are geodesics. Figure \ref{emesh} illustrates this relative geometry, which is to be compared with the trivial background structure typically assumed for approximately flat space-time. This figure is intended to serve as a simple, explicit example of relative geometry, to make it easier to grasp that new concept.

However, figure \ref{emesh} clearly is not the full relative geometry even for a single slit: the true $g_{ij}(x,y)$ needs to be both smooth and include edge diffraction. In the case of the single slit, these corrections are not difficult to picture. The edge modulation (discussed in the next section) can be added to the foci, and for smoothness a local modulation along $a=0$ (of the elliptic frame) can be introduced. It need not change the qualitative features of the relative geometry overly much.

The double-slit relative geometry is more complicated. A brief speculation on how one might solve for that geometry can be found in \S\ref{s.infoflow}, which indicates a nested set-up with modifications mostly around the foci, provided a clear separation of scales. A key question is how the different single- and double-slit intensity distributions, with the latter periodic in $\lambda/D$ throughout and the former with a centralised peak of width $2\lambda/D$, could possibly be a product of the same type of geometry, that of figure \ref{emesh}. However, there is a decided difference in that the path through a single slit is not centred through a foci, but along the line in-between. Likely the experienced geometry can account for the dissimilarity of the single- and the double-slit diffraction periodicities. A conclusive statement would require a more thorough analysis of how the particle paths evolve.

At the present level of analysis, the relative geometry of figure \ref{emesh} (with suitable modifications) implies little new for the paths of particles passing through. The geometry scales linearly with $D$, which fits with a a scale invariance of $\lambda/D$ and implies that quantum resonances occur at multiples of $C \lambda/D$ for some constant $C$. The symmetry sets the main peak to be at no angular deviation, and the path deviations are restricted to occur within some distance $\sim D$ of the foci, since the relative geometry quickly approaches the reference frame. In addition, slits and gratings would give similar results, with particle paths going through the half-plane relative geometry twice in both settings.

The conclusion from this merely is that it likely is possible to fashion a structure within flat space-time and endow particle interactions with overall properties in relation to it, with an end result in rough agreement with light diffraction. The slit illustration is an explicit example of how that structure would appear. We now turn to an example that straightforwardly gives a smooth relative geometry, and has implications for the quantum interactions.

\subsubsection{Edge diffraction vs Gaussian apertures}\label{s.Gauss}
The symmetry of a sharp edge is not obvious. Assuming that relative geometry occurs at \emph{changes} in information connectivity, a circular symmetric, smooth modulation with an exponential convergence to the reference frame is implied. The simplest example is a Gaussian, but the deduction of that shape can be made more precise. The inverse of what relative geometry a sharp edge causes is when the trivial reference frame is valid across an aperture. Such apertures give intensity distributions directly corresponding to the transmission profile of the aperture, which holds precisely for Gaussian profiles. Hence the relative geometry around a sharp edge indeed should be described by a Gaussian modulation.

In polar coordinates $(r,\theta)$ set relative to the reference frame and centred on the tip of the edge, we then have geodesics curving around the edge with
\be
dl^2_{d=2}=d\xi^2+d\eta^2\,,\qquad\left\{\begin{array}{l}\xi=\left(1-e^{-\frac{r^2}{2\sigma^2}}\right)r\cos\theta\\\eta=\left(1-e^{-\frac{r^2}{2\sigma^2}}\right)r\sin\theta \end{array}\right.\,,
\ee
where the standard deviation $\sigma$ represents a length scale of the underlying structure.

\begin{figure}[tbp]
\begin{center}
\includegraphics[scale=0.67]{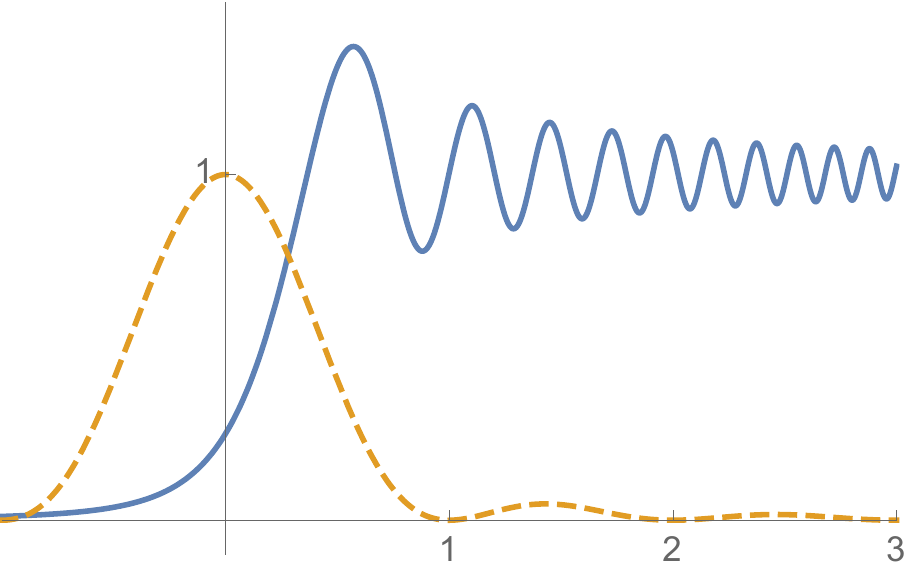}
\end{center}
\vspace{-0.5cm}
\caption{The qualitative features of single-slit and edge diffraction (intensity vs angular deviation). The only comparison intended is a note on the periodicity, which is linear for a slit but approaches quadratic for a sharp edge.}\label{sve}
\end{figure}

The above proposed relative geometry has definite similarities to actual edge diffraction. The Fresnel integrals used for describing the diffraction pattern are reminiscent of $e^{i r^2}$, and for quantum effects in terms of path deviations, the Gaussian profile suggest a periodicity in something close to or approaching $x^2$ rather than $x$, as is the case for the slit periodicities of $n \times \lambda/D$ with the linear exponent in \eqref{eq.eflat}. The $\sim x^2$ periodicity is also a property of edge diffraction, depicted in figure \ref{sve}. While far from conclusive, these similarities are non-trivial and in line with a correlation between the diffraction pattern and the relative geometry. 

\subsubsection{The impact of information connectivity on flat space-time}\label{s.infoflow}
\begin{figure}[tb]
\begin{center}
\begin{tikzpicture}[scale=1.1]
\draw (0,0) -- (1.5,0) (2.5,0) -- (4,0) (2,-0.2) -- (2,1.4) (1.75,-0.2) -- (1.75,1.4) (2.25,-0.2) -- (2.25,1.4);
\draw [->] (1.5,0.31) arc (30:90:20pt);
\draw [->] (2.5,0.31) arc (150:90:20pt);
\filldraw (0.5,0.3) circle (0.5pt) (0.9,0.3) circle (0.5pt) (0.5,0.7) circle (0.5pt) (1.3,1.1) circle (0.5pt) (0.5,1.1) circle (0.5pt) (1.3,0.3) circle (0.5pt) (0.9,1.1) circle (0.5pt) (1.3,0.7) circle (0.5pt);
\filldraw (3.4,0.3) circle (0.5pt) (3.1,0.3) circle (0.5pt) (3.4,0.7) circle (0.5pt) (2.7,1.1) circle (0.5pt) (3.4,1.1) circle (0.5pt) (2.7,0.3) circle (0.5pt) (3.1,1.1) circle (0.5pt) (2.7,0.7) circle (0.5pt);
\end{tikzpicture}
\hspace{1cm}
\includegraphics[scale=0.4]{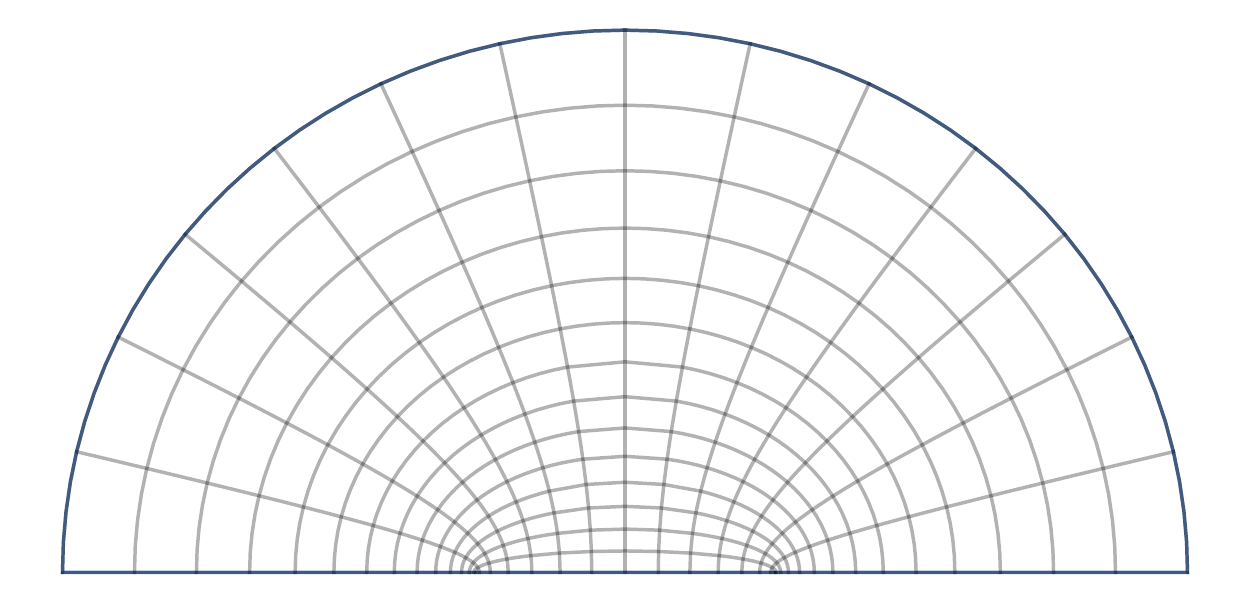}
\vspace{-0.5cm}
\end{center}
\caption{Illustration of (\emph{left}) how a higher number of degrees of freedom are available on each side of a constriction, to a flow passing through, and (\emph{right}) the suggested elliptic (one-sided) solution to the geodesics, disregarding smoothness and edge effects at the connecting line.}\label{dofs}
\end{figure}
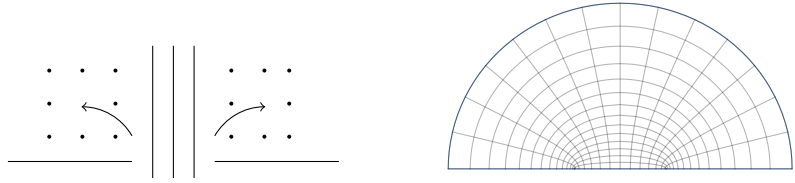

An emergence of relative geometry from information connectivity can be examined at the macroscopic level, before a closer look at what would be required of the quantum interactions. Here, the space-time description simply is that of general relativity, but with the assumption of an underlying structure in information exchange between particles (quantum fluctuations): to what degree they are entangled and how quickly a deviation is communicated from one point to another. The simplest illustration of the subsequent effect is in flat space-time, where the density of quantum fluctuations can be assumed to be constant compared to the vacuum reference frame, and the intuitive picture is that of figure \ref{dofs}.
At any constriction formed by matter, information will be more quickly communicated through the empty space, i.e. the openings or connecting regions. For simplicity we assume the boundaries set by matter to be comparatively insulating, in the same sense that they block photons, which ought to be a good first order approximation. When a connecting region opens up, the rate of interaction is diluted among the increased amount of available points of interaction. The quantum fluctuations cannot be equally tightly bound together throughout. The result is a reference frame for length and directions which describes flat space-time, but carries relative geometry compared to a configuration without the obstacle in question. Effectively, it should be possible to think of the equilibrating process as a flow of information, although the static case does not contain an actual flow through the space-time, but represents how the structure is upheld by continuous equilibration of connections between the particles. This is our qualitative interpretation of the underlying process. The subsequently required, suitable interactions at the quantum level are discussed in \S\ref{s.qnat}.

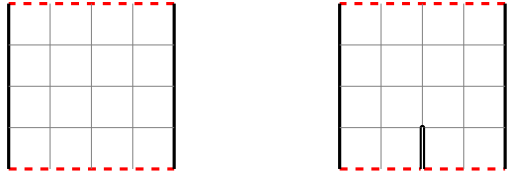
\begin{figure}[t]
\begin{center}
\begin{tikzpicture}[scale=1.1]
\draw[red,dashed,very thick] (0,0) -- (2,0) (2,2) -- (0,2);
\draw[very thick] (2,0) -- (2,2) (0,0) -- (0,2);
\draw[gray, very thin] (0,1) --(2,1) (0,0.5) --(2,0.5) (0,1.5) --(2,1.5);
\draw[gray, very thin] (1,0) -- (1,2) (1.5,0) -- (1.5,2) (0.5,0) -- (0.5,2);
\draw[red,dashed,very thick] (4,0) --(4.98,0) (5.02,0) -- (6,0) (6,2) -- (4,2);
\draw[very thick] (6,0) -- (6,2) (4,0) -- (4,2);
\draw[gray, very thin] (4,1) --(6,1) (4,0.5) --(4.98,0.5) (5.02,0.5) --(6,0.5) (4,1.5) --(6,1.5);
\draw[gray, very thin] (5,0.52) -- (5,2) (5.5,0) -- (5.5,2) (4.5,0) -- (4.5,2);
\draw[thick] (4.98,0) -- (4.98,0.5) (5.02,0) -- (5.02,0.5);
\draw[thick] (5.02,0.5) arc (0:180:0.2mm);
\end{tikzpicture}
\end{center}
\vspace{-0.5cm}
\caption{Half-planes of (\emph{left}) the single-, and (\emph{right}) the double-slit geometries, recast through diffeomorphisms into a linear setting with insulating vs connecting boundaries at right angles. This gives trivial geodesics and a tentative way of identifying more intricate relative geometries.}\label{bound}
\end{figure}

When identifying a relative geometry, i.e. the macroscopic structure, a useful approach seems to be to use diffeomorphism invariance to render the geometry into straight lines at right angles, with parallel lines denoting the same type of boundary condition (connecting vs insulating), as illustrated in figure \ref{bound}. In these settings the geodesics are trivial (up to edge effects and smoothness across connecting lines) and the diffeomorphism to get to the linear setting is the key to the relative geometry. The linear configuration is easily obtained for each half-plane of the single slit in terms of elliptic coordinates, but to find the appropriate diffeomorphism for e.g. the double slit is more difficult.

In considering an information origin of space-time, the flat space-time structure turns into a boundary value problem, modulus Gaussian profiles. A resemblance to temperature in figure \ref{bound} is easy to see in solutions to the heat equation in the absence of sources ($\nabla^2T=0$), but where temperature originates in linear interactions, the information structure is characterized by something else. With energy $E\propto r^{-2}$ (instead of $E\propto T$), curvature definitely should represent non-linear dynamics at the quantum level. The qualitative macroscopic picture is of a tension structure. The flatness condition, in $2d$ simply \eqref{2flat}, is diffeomorphism invariant and only compatible with linear interactions when the boundary conditions are linearized, as in figure \ref{bound}. Meanwhile, the diffeomorphism invariance amounts to a purely relative description, a self-consistent theory without external points of reference (parameterization invariant).

\section{The quantum nature of the interactions} \label{s.qnat}
At the quantum level, the ansatz of an origin of space-time in information exchange means that we model a line element $ds^2$ as arising from interactions between quantum particles: mostly quantum fluctuations in the vacuum, but any other particles as well. With a qualitative separation $ds^2=-c^2dt^2+dl^2$, the space-like distance $dl$ can be substituted with an effective interaction rate ($f$) through assuming information to be communicated at the speed of light, $c$: $dl=c/f$. $f$ is in turn given by an average over the individual interaction rates between quantum particles in a region. $dt$ is also connected to an interaction rate: that of the total interaction rate for each quantum particle, which provides a relative measure of the passage of time for each particle. This way of modelling time is intuitive from the nature of atom clocks and effects on them from gravity.\footnote{A model where regions near matter are characterized by less frequent interactions (by gravity) also justifies the assumption in \S\ref{s.flat} that geometry in approximately flat space-times can be modelled as insulating boundaries to information flow. There, it is reasonable to treat lines/areas of slower information exchange as insulating, in some simplifying limit.}

As a starting point, the points of reference can be thought of as on a lattice, and an equilibrating process is assumed to be present so that differences in interaction rates are suppressed. In the classical, static and flat model analysed in \S\ref{s.flat}, effects of quantum fluctuations, gravity, time-dependence and diffeomorphisms changing the weight of $dt$ are disregarded. The only relevant interaction rate for a quantum particle then is how frequently it interacts with different particles. By causality the (static) interactions must be consistent with a lattice distribution, so that interaction between two particles, where one is connected to a third party, implies direct interactions between all of the tree particles in consistency with the mathematical definition of a neighbourhood.

However, while a lattice picture is helpful, it is both misleading and strictly not necessary. To begin with, there is a redundant degree of freedom in the choice of coordinate frame, appearing to give rise to diffeomorphism invariance, since all that matters is the \emph{relative} total rate and connections between the quantum particles rather than the way those are presented. In addition, in the flat case it is tempting to give preference to the equidistant lattice with identical fall-off in interaction rate, corresponding to a Minkowski metric. This picture is misleading since the quantum particles representing quantum fluctuations neither appear at set points nor last indefinitely, and the density (of quantum fluctuations) is expected to be translation invariant in the vacuum reference frame rather than in the diagonalised representation, if it fills any physical function at all. In the presently analysed hypothesis of space-time emergence, the equivalence class that is diffeomorphism invariance makes interaction rate the only invariant, definitely physical property.

In a classical limit, a space-time originating in interactions must have the properties of a tension structure, built on non-linear interactions. The boundary conditions alone need to set the static, flat configuration, as illustrated in \S\ref{s.infoflow}. At the quantum level, the interactions must be compatible with giving rise to such a macroscopic, flat structure, and with the EFE in general, in a limit where quantum fluctuations are suppressed to the point when they can be disregarded. These classical theories should be good descriptions across macroscopic times and distances (where quantum distances are at the Planck scale) in regions where the lifetime of the quantum fluctuations $(\tau_f)$ is large enough for the structure to equilibrate $(\tau_f\gg\tau_{eq})$.

\subsection{Restrictions from flat, static space-time}\label{s.qrest}
It is unclear if restrictions on the quantum interactions can be deduced directly from the effective tension structure discussed in \S\ref{s.infoflow}. A complicating circumstance is that the setting is macroscopic.

As to the profile of the quantum interaction rates (in equilibrium), the fall-off of the profile is implied by several properties of the static, flat space-time. (1) As analysed in \S\ref{s.Gauss}, the fall-off of relative geometry away from geometry without an imposed scale is Gaussian, and relative geometry is sensitive to boundary conditions modulus Gaussian profiles. (2) By smoothness of $g_{\mu\nu}$, the functional basis ought to be Gaussian. (3) Gaussians interactions give rise to Euclidean geometry. Relations such as Pythagoras' theorem (and relations with angles in general) can be recovered from that consecutive interactions, by necessity products, add up in terms of lengths. The interactions e.g. must be described by some function $f$ fulfilling
\be\label{eq.abc}
f(a)f(b)=f(c) \quad \text{as in}\quad e^{a^2}e^{b^2}=e^{c^2}
\ee
for some special directions $(\vec{a}\perp\vec{b})$, presumably in a unitary representation. All of this implies a consistent ansatz to be that each quantum particle is associated with a Gaussian interaction rate profile,
\be\label{eq.Gaussian}
f_t\times\frac{1}{\prod_i\sigma_i}e^{-\sum_i\frac{x_i^2}{2\sigma_i^2}}\,,\quad i\in\{1,2,\ldots,d\}\,.
\ee
The interaction rate profile is interpretable from a classical perspective as a statistical distribution for how often a particle \emph{initiates} interaction with particles in its vicinity (not to be confused with interactions initiated by other particles). The standard deviations $\sigma_i$ set a reference for length through describing how quickly the interaction frequency decreases away from the one-to-one correlation at $x=0$. We will assume this reference length to be naturally unitary in vacuum: $\sigma_{0,i}=1$. $f_t$ gives a measure of the individual, total interaction rate, and we will let $f_t\leq f_{t,0}=1$. In total, the interaction rate profile is quite similar to a wave function, if interaction is considered as some type of collision between particles while particle position has a distribution (with a standard deviation) due to complementarity. However, our consideration rather is prior to a concept of distance and position, and concerns interaction rates that in some sense describe probabilities of interaction.

In this picture, deviations from the vacuum values should be due to acceleration or fall under the equivalence class of diffeomorphisms. A deviation from the Gaussian profile itself is expected in a pre-equilibration state of the quantum fluctuations, and perhaps even due to acceleration. The profile is an idealization in the sense that its spread must be limited by causality. Note that the suggested interaction profile is derived only from classical structures: the static, flat scenario analysed in \S\ref{s.flat} and properties of Euclidean space and the metric. While these properties seem to be internally consistent and in agreement with an information origin of space-time, they say nothing of the nature of quantum fluctuations, or of restrictions from properties of curvature on a macroscopic scale. For example, in the static, flat case all quantum particles have $f_t\equiv1$ and are identical save for the perceived distance between them, and so the analysis only concerns interaction rates between points, as in $\sigma$, while omitting relations concerning $f_t$. It is a decoupled set-up, in need of more analysis to be made precise.

Although the analysis decouples properties related to time from those of space and only probe the latter, a statistical average is implied:
\be
\textstyle ds^2\sim\left\langle-c^2f_t^2({\bf x})dt^2+\sum_i\sigma_i^{-2}({\bf x})dx_i^2\right\rangle
\ee
in the coordinates of a flat, vacuum reference frame, where the configuration ($\sigma_i$) needs to be shaped by some flatness condition. Invariance under a change of inertial frame does not give more details on the static, flat interactions --- what goes beyond a rephrasing of how the interactions appear rather falls under effects of acceleration (such as a slower passage of time) and quantum effects at speeds $v\sim c$. 

A final comment that can be made about the interactions based on the static and flat analysis is that the presence of a set $\{\sigma_i\}$ implies multiple separate and orthogonal interaction channels for each particle, as is characteristic for spin 1/2 entanglement. Such a basis has the potential of capturing effects of polarization, typical for interference of light, with separate, co-existing reference structures. In addition, this would give $[S_x,S'_y]\propto e^{-\sum_i({\bf x}_{i}-{\bf x}'_{i})^2/(2\sigma_i^2)}$ for two separate particles, provided that the entanglement is proportional to the interaction. In this sense, interaction rates possibly may be interpreted in terms of entanglement entropy.

\subsection{Comments on effects of curvature}
The hypothesis of an interaction origin of space-time needs an analysis with respect to curvature and effects at the quantum level, beyond the scope of this text. The flat, static case discussed above must have additional restrictions from curvature/acceleration and time-dependence, including quantum fluctuations. Some immediate observations can however be made. The effect of acceleration on local, relative time is visible in the weights of $dt^2$ in terms of the Lorentz factor and the Schwarzschild metric (with $c=1$)
\be
\left(1-v^2\right)dt^2\,, \quad \left(1-\frac{2MG}{r}\right)dt^2\,,
\ee
relatable to kinetic $(mv^2/2)$ and potential energy $(mMG/r)$, indicating that the weights should be possible to attribute to the force a (virtual) particle has been subjected to. One question is if this can be connected to the quantum physics of the interaction rates, e.g. in terms of complementarity of $(E,t)$. Invariance of $\textstyle f_t/\prod_i\sigma_i$ under a change of energy would set length contraction and the inverse of time dilation (i.e. the frequency counterpart) to be identical. 

A more direct feature is that there for gravity is a non-linear equilibrium tension configuration, visible e.g. in the $r^{-1}$ of the $g_{tt}$ of the Schwarzschild metric, which the interactions must capture on a macroscopic scale, in addition to compatibility with the EFE and properties of curved space. The non-linear profile suggests a dependence on both $f_t$ and $\{\sigma_i\}$ in the near vicinity, which would be compatible with changes propagating through space-time as waves, in a description where a local equilibration of fast modes (non-classical) has been projected out. Considering this and the previous analysis of flat space-time, it would not be surprising if general relativity can be modelled consistently as emergent from information exchange. The question is if the quantum effects are realistic.

Another aspect of the quantum particle scheme is that quantum fluctuations and particles (separate from the vacuum by a longer lifetime) are put on an equal footing in terms of upholding the interactions giving rise to space-time --- with the difference that non-fluctuations remain present, retain their properties and ought to have a set total interaction rate relative to the surroundings, so that they define the structure rather than carry it. In a full consistency check of the hypothesis, it is necessary to examine if this equal treatment of quantum particles can be compatible with how the stress-energy tensor shows in the EFE.

Even with a full analysis of what general relativity would specify for the quantum interactions (if compatible) most of the quantum features would still be undetermined. For a full picture not only the quantum interactions themselves would have to be identified, but also the manner in which they equilibrate: how changes propagate etc. The equilibration protocol at the quantum level would be especially relevant for the quantum properties of gravity, and would define the physics near event horizons. In the present hypothesis, event horizons occur at $\tau_f\sim\tau_{eq}$, where the frequency of interaction between quantum fluctuations is not large enough to allow information (including light) to propagate away and have the connections equilibrate to the classical limit, before the fluctuations cease to exist. Here, quantum fluctuations of space-time (not captured by the EFE) would dominate.

A breakdown of the effective theory described by general relativity close to event horizons would be of central interest in a final analysis, representing the main reason to consider an interaction origin of space-time to begin with. Based on the assumption of an origin of propagation in information exchange between quantum particles, a region with $\tau_f\sim\tau_{eq}$ can be expected to be characterized by a randomization of propagation. This might give rise to a shell-like structure around black holes, with a sharp fall-off of probability of particle occurrence in the radial direction, and a random output of `interior' particles as they reconnect with the equilibrated space-time. This is however pure speculation. It is difficult to say what could be the result of a scheme where the existence of a particle would have as much impact as anything in the surroundings on the near vicinity `space-time'. A random walk scenario with a shell-like structure (where the interior of a black hole is left as outside of/a sparse area of space-time) however has similarities to the fuzzball proposal \cite{Mathur:2005zp} and suggestions that the event horizon of a black hole represent a quantum phase transition \cite{Chapline:2000en}.

\section{Summary}
Quantizations of gravity typically build on that space-time constitutes a background insensitive to changes in physics above the Planck scale. This assumption of insensitivity is especially important at event horizons, where physics crucial to quantum gravity is expected and light propagation is disrupted. Motivated by the elusive nature of quantum gravity, its connection to entanglement (i.e. interactions, if dynamic), and to exhaust all possibilities, we have looked at what would come out of a scenario where space-time is not insensitive to disrupted propagation of information, such as light. We have initiated a study of a hypothesis in which classical space-time arises as an effective tension structure for how information propagates, upheld in the vacuum by information exchange (interactions) between quantum fluctuations. In this picture, a line element would arise as a statistical average of how frequently particles interact, through an individual rate $dt\sim 1/f_t$ and interconnecting rates $dl\sim c/f$.

We have analysed if this type of quantization can be modelled in a way consistent with emergence of flat space-time, and illustrated the effects it would have on the emergent, classical structure as well as what would be required of the quantum interactions. In addition to curvature, a space-time originating in information exchange would be shaped by how matter provides boundary conditions to propagation of information, a scenario which could be compatible with wave-particle duality in particle diffraction. However, our only non-trivial result is that equilibrated interactions must have a Gaussian fall-off to capture the relations of Euclidean geometry, which coincides with trivial particle diffraction for Gaussian apertures. More thorough analyses of quantum effects, and curvature, are necessary to determine whether the hypothesis is realistic or not. Apart from quantum properties of the space-time structure, a second type of quantum effect is present through how particles interact with the space-time. If the interference effects of particle diffraction cannot fit with the scenario of relative geometry, the hypothesis fails.

Our focus has been on a rough, qualitative analysis of what would be required for consistency rather than on counterexamples, which constitutes a limited scrutiny of the hypothesis, but we feel enough internal consistency is observed to warrant further interest in this ansatz for quantizing space-time.

\section*{Acknowledgements}
This work is supported by the Swedish Research Council grant 2017-00328.

\providecommand{\href}[2]{#2}\begingroup\raggedright\endgroup
\end{document}